\documentclass[useAMS,usenatbib]{mn2e}

\usepackage{graphicx}

\title[Parsec-scale jet precession in BL Lacertae]{Parsec-scale jet precession in BL Lacertae (2200+420)}
\author[A. Caproni, Z. Abraham and H. Monteiro]{A. Caproni$^{1}$\thanks{E-mail:
    anderson.caproni@cruzeirodosul.edu.br}, Z. Abraham$^{2}$ and H. Monteiro$^{3}$\\
  $^{1}$N\'ucleo de Astrof\'\i sica Te\'orica, Universidade Cruzeiro do Sul, R. Galv\~ao Bueno 868, Liberdade, 01506-000, S\~ao Paulo, SP, Brazil\\
  $^{2}$Instituto de Astronomia, Geof\'\i sica e Ci\^encias Atmosf\'ericas, Universidade de S\~ao Paulo, R. do Mat\~ao 1226, Cidade Universit\'aria,\\
  05508-900, S\~ao Paulo, SP, Brazil\\
  $^{3}$ UNIFEI, Instituto de Ci\^encias Exatas, Universidade Federal
  de Itajub\'a, Av. BPS 1303, Pinheirinho, 37500-903, Itajub\'a, MG,
  Brazil}

\begin{document}

\date{Accepted 1988 December 15. Received 1988 December 14; in original form 1988 October 11}

\pagerange{\pageref{firstpage}--\pageref{lastpage}} \pubyear{2002}

\maketitle

\label{firstpage}

\begin{abstract}
BL Lacertae is the prototype of the BL Lac class of active galactic nuclei, exhibiting intensive activity on parsec (pc) scales, such as intense core variability and multiple ejections of jet components. In particular, in previous works the existence of precession motions in the pc-scale jet of BL Lacertae has been suggested. In this work we revisit this issue, investigating temporal changes of the observed right ascension and declination offsets of the jet knots in terms of our relativistic jet-precession model. The seven free parameters of our precession model were optimized via a heuristic cross-entropy method, comparing the projected precession helix with the positions of the jet components on the plane of the sky and imposing constraints on their maximum and minimum superluminal velocities. Our optimized best model is compatible with a jet having a bulk velocity of 0.9824$c$, which is precessing with a period of about 12.1 yr in the observer's reference frame and changing its orientation in relation to the line of sight between 4$\degr$ and 5$\degr$, approximately. Assuming that the jet precession has its origin in a supermassive binary black hole system, we show that the 2.3-yr periodic variation in the structural position angle of the very-long-baseline interferometry (VLBI) core of BL Lacertae reported by Stirling et al. is compatible with a nutation phenomenon if the secondary black hole has a mass higher than about six times that of the primary black hole.
\end{abstract}

\begin{keywords}
galaxies: active -- BL Lacertae objects: individual: (BL Lacertae) -- galaxies: jets -- radio continuum: galaxies. 
\end{keywords}

\section{Introduction}

BL Lacertae or 2200+420 (z=0.0686; \citealt{ver95}), the prototype of the BL Lac class of active galactic nuclei, is hosted by an elliptical galaxy, composed mainly by a stellar population of about 0.7 Gyr \citep{hyv07}. As usual in BL Lac type objects, its nuclear region exhibits strong continuum variability on different time-scales, from days to years, over the whole  electromagnetic spectrum.

High-resolution interferometric images at radio wavelengths show the presence of a compact core and a diffuse halo-like source at arc-second scales \citep{anto86}. At an intermediate resolution ($\sim$ 10 mas), BL Lacertae shows a core-jet structure, with components following different trajectories on the plane of the sky \citep{pol95}. Similar behavior is seen at mas resolution, which has been attributed to either helical instabilities (e.g., \citealt{tat98,den00}) or jet inlet precession \citep{sti03, tate09}. Concerning jet precession, \citet{sti03} proposed a period of 2.29 years based on the periodic variations in the polarization position angle at 1 mm, and in the direction of the innermost radio core component at 43 GHz, which has been matter of criticism in some recent works (e.g., \citealt{mude05}). \citet{tate09} proposed an alternative precession model in which the BL Lacertae parsec-scale jet changes its orientation in a period of about 26 yr. Such claim was based on analyses of maps of BL Lacertae at 8 and 15 GHz in a super-resolution mode.

In this work, we reanalyse the jet-precession proposition in terms of our ballistic jet precession model, described in Section 2. In Section 3, we introduce the cross-entropy (CE) global optimization technique in the context of our precession model. In Section 4, we show the general results from the application of our CE jet-precession model to BL Lacertae, as well as the observational constraints used in this work. We explore in Section 5 the possible consequences of the underlying jet intensity on the radio and optical historical light curves. In Section 6 we study the viability of a supermassive binary black hole system in the nuclear region of BL Lacertae in producing
the inferred jet precession rate, as well as a nutation motion with period of 2.3 yr. Finally, general conclusions are presented in Section 7.

We will assume throughout this work a $\Lambda$CDM cosmology with $H_0=71$ km s$^{-1}$ Mpc$^{-1}$, $\Omega_\rmn{M}=0.27$, and $\Omega_\rmn{\Lambda}=0.73$, which implies 1 mas = 1.296 pc and 1 mas yr$^{-1}$ = 4.516$c$ for BL Lacertae.

\section{Ballistic Jet Precession Model}

Let us consider a relativistic jet receding from the core with a constant bulk velocity $\beta$ (in units of the speed of light $c$) that precesses around a fixed axis (see Fig. 1 in \citealt{cap09} for a schematic representation of this). Precession makes the jet inlet direction vary with time with a precession period $P_\rmn{prec,s}$ measured in the source reference 
frame, producing a cone with semi-aperture angle $\varphi_0$. The precession phase 
$\omega_\rmn{s}\Delta t_\rmn{s}=2\pi(t_\rmn{s}-t_\rmn{0,s})/P_\rmn{prec,s}=2\pi(\tau_\rmn{s}-\tau_\rmn{0,s})$ is chosen arbitrarily to be 
zero on the $y_\rmn{s}z_\rmn{s}$-plane at $\tau_\rmn{s}=\tau_\rmn{0,s}$ \citep{cap09}. From these  definitions, the 
instantaneous jet inlet direction is given in terms of a unit vector with Cartesian components:

 \[
   e_\rmn{x,s}(\tau_\rmn{s})=\sin\varphi_0\sin[\iota2\pi(\tau_\rmn{s}-\tau_\rmn{0,s})],
 \]
 \[
   e_\rmn{y,s}(\tau_\rmn{s})=\sin\varphi_0\cos[\iota2\pi(\tau_\rmn{s}-\tau_\rmn{0,s})],
 \]
 \[
   e_\rmn{z,s}(\tau_\rmn{s})=\cos\varphi_0,  
 \]
where $\iota$ gives the sense of precession, $\iota=1$ for clockwise sense and $\iota=-1$ for counterclockwise precession.

Introducing the parameters $\phi_0$, the angle between the precession cone axis and the line of sight, and  $\eta_0$, the position angle of the axis on the plane of the sky (positive from north to east), we have (e.g., \citealt{caab04a,caab04b,cap09}):

\begin{equation}
  e_\rmn{x,obs}(\tau_\rmn{s})=A(\tau_\rmn{s})\cos\eta_0-e_\rmn{y,s}(\tau_\rmn{s})\sin\eta_0,
\end{equation}

\begin{equation}
  e_\rmn{y,obs}(\tau_\rmn{s})=A(\tau_\rmn{s})\sin\eta_0+e_\rmn{y,s}(\tau_\rmn{s})\cos\eta_0,
\end{equation}

\begin{equation}
  e_\rmn{z,obs}(\tau_\rmn{s})=-e_\rmn{x,s}(\tau_\rmn{s})\sin\phi_0+e_\rmn{z,s}(\tau_\rmn{s})\cos\phi_0,
\end{equation}
 where:
\begin{equation}
  A(\tau_\rmn{s})=e_\rmn{x,s}(\tau_\rmn{s})\cos\phi_0+e_\rmn{z,s}(\tau_\rmn{s})\sin\phi_0.
\end{equation}

The instantaneous angle between the jet and the line of sight $\phi$ is calculated from:

\begin{equation}
  \phi(\tau_\rmn{s})=\arccos[e_\rmn{z,obs}(\tau_\rmn{s})],
\end{equation}
 while the position angle of the jet on the plane of the sky $\eta$ is obtained from:
 
\begin{equation}
  \eta(\tau_\rmn{s})=\arctan\left[\frac{e_\rmn{y,obs}(\tau_\rmn{s})}{e_\rmn{x,obs}(\tau_\rmn{s})}\right].
\end{equation}

The observed jet velocity $\beta_\rmn{obs}(\tau_\rmn{s})$ is:

\begin{equation}
  \beta_\rmn{obs}(\tau_\rmn{s})=\gamma\beta\delta(\tau_\rmn{s})\sin\phi(\tau_\rmn{s}),
\end{equation}
 where the jet Lorentz factor $\gamma$ is  
 
\begin{equation}
  \gamma=\left(1-\beta^2\right)^{-1/2},
\end{equation}
 and the jet Doppler factor $\delta$ is: 

 \begin{equation}
  \delta(\tau_\rmn{s})=\gamma^{-1}\left[1-\beta\cos\phi(\tau_\rmn{s})\right]^{-1}.
\end{equation}

In order to compare predictions from the precession model with observational data, it is necessary to transform the elapsed time measured in the source's reference frame $dt_\rmn{s}$ to the time interval in the observer's framework $dt_\rmn{obs}$. Following \citet{gow82} and \citet{cap09}, we can write:

\begin{equation}   
\frac{\Delta t_\rmn{obs}}{P_\rmn{prec,obs}}=\frac{\int_{0}^{\Delta\tau_\rmn{s}}\delta^{-1}(\tau)d\tau}{\int_0^1\delta^{-1}(\tau)d\tau},
\end{equation}
 where $\Delta t_\rmn{obs}=\left(t_\rmn{obs}-t_\rmn{0,obs}\right)$ and $\Delta\tau_\rmn{s}=\tau_\rmn{s}-\tau_\rmn{0,s}$.

The relation between the precession period in the source's framework and that measured by the observer $P_\rmn{prec,obs}$ is given as:

\begin{equation}     
   P_\rmn{prec,s}=\frac{\gamma}{(1+z)}\frac{P_\rmn{prec,obs}}{\int_0^1\delta^{-1}(\tau)d\tau},
\end{equation}
where $z$ is the redshift of the source.

A fluid element of the jet, ejected at time $t_\rmn{obs,ej}$ from a jet inlet region that is precessing according to our model, will be observed at time $t_{\rm obs}$ with  right ascension and declination offsets from the core $\Delta\alpha_\rmn{mod}$ and $\Delta\delta_\rmn{mod}$ respectively, given by: 

\begin{equation}     
  \Delta\alpha_\rmn{mod}(t_{\rm obs}) = \left(t_{\rm obs}-t_\rmn{obs,ej}\right)\mu(t_\rmn{obs,ej})\sin[\eta(t_\rmn{obs,ej})],
\end{equation}

\begin{equation}     
  \Delta\delta_\rmn{mod}(t_{\rm obs}) = \left(t_{\rm obs}-t_\rmn{obs,ej}\right)\mu(t_\rmn{obs,ej})\cos[\eta(t_\rmn{obs,ej})],
\end{equation}
where $\mu$ is the proper motion of the fluid element predicted by our jet-precession model assuming outward motion at a constant speed. 

Considering all the possible values of $t_{\rm obs,ej}<t_{\rm obs}$ we obtain a snapshot of the model jet at time $t_{\rm obs}$, projected on the plane of the sky. The real jet, as observed with VLBI techniques,  is not continuous but formed by discrete components, ejected at unknown epochs with unknown velocities. To avoid incorrect identifications of the same component in maps obtained at different epochs (which could led to incorrect determination of $t_\rmn{obs,ej}$ and $\mu(t_\rmn{obs,ej})$ for this component), we will use only the projected position of the components in the plane of the sky for each epoch in which a map is available, and compare them with the positions predicted by the model.

In summary, our jet precession model has seven free parameters ($P_\rmn{prec,obs}$, $\iota$, $\gamma$, $\eta_0$, $\phi_0$, $\varphi_0$ and $\tau_\rmn{0,s}$), which are determined, via the CE technique as will be discussed in the next Section.

\section{The cross-entropy precession method}

\subsection{General overview}

The CE method was originally employed in the optimization of complex computer simulation models involving rare events simulations \citep{rubi97}, having been modified by \citet{rubi99} to deal with continuous multi-extremal and discrete combinatorial optimization problems. Its theoretical asymptotic convergence in such situations has been demonstrated by \citet{marg04}, while \citet{kro06} studied the efficiency of the CE method in solving continuous multi-extremal optimization problems. Some examples of robustness of the CE method in several situations are listed in \citet{deb05}. 

CE optimization involves basically random generation of the initial parameter sample (obeying some predefined criteria) and selection of the best samples based on some mathematical criterion. Subsequent random generation of updated parameter samples from the previous best candidates are performed iteration by iteration until a pre-specified stopping criterion is fulfilled.

Application of the CE method in astrophysical contexts can be found in \citet{cap09}, \citet{mon10}, \citet{cap11} and \citet{modi11}.

\subsection{Our CE algorithm for jet precession modelling}

Following \citet{cap09}, let us suppose that we wish to study a set of $N_\rmn{d}$ observational data in terms of an analytical model characterized by $N_\rmn{p}$ parameters $p_1, p_2, ..., p_{N_\rmn{p}}$. In the case of jet precession modelling, the observational data correspond to right ascension and declination offsets, $\Delta\alpha$ and $\Delta\delta$ respectively, for the $N_\rmn{d}$ knots. As mentioned before, our precession model is defined by the free parameters $\beta$ (or $\gamma$), $\eta_0$, $\phi_0$, $\varphi_0$, $\tau_\rmn{0,s}$, $P_\rmn{prec,obs}$ (or $P_\rmn{prec,s}$) and $\iota$, i.e., $N_\rmn{p}=7$. In this work, we adopted the same procedure suggested by \citet{cap09}, in which for a given (fixed) value of $P_\rmn{prec,obs}$ and $\iota$, the remaining five parameters are CE optimized. Therefore, it is necessary to test different values for the precession period, as well as distinct senses of precession in order to obtain the best set of model parameters.

The main goal of the CE continuous multi-extremal optimization method is to find the set of parameters ${\bf x}^*=(p^*_1,p^*_2,...,p^*_{Np})$ for which the model provides the best description of the data \citep{rubi99,kro06}. It is performed generating randomly $N$ independent sets of model parameters ${\bf X}=({\bf x}_1,{\bf x}_2,...,{\bf x}_N)$, where ${\bf x}_i=(p_{1i},p_{2i},...,p_{N{\rmn{p}i}})$, and minimizing a merit function $S({\bf x})$ used to transmit the quality of the fit during the run process. If the convergence to the exact solution is achieved then $S({\bf x}^*)\rightarrow 0$.

In order to find the optimal solution from CE optimization, we start by defining the parameter range in which the algorithm will search for the best candidates: $p^\rmn{min}_j\leq p_j(k) \leq p^\rmn{max}_j$, where $k$ represents the iteration number. Introducing $\bar{p}_j(0)=(p^\rmn{min}_j+p^\rmn{max}_j)/2$ and $\sigma_j(0)=(p^\rmn{max}_j-p^\rmn{min}_j)/2$, we can compute ${\bf X}(0)$ from:

\begin{equation}
  X_{ij}(0)=\bar{p}_j(0)+\sigma_j(0) G_{ij},
\end{equation}
where $G_{ij}$ is an $N\times N_\rmn{p}$ matrix with random numbers generated from a zero-mean normal distribution with standard deviation of unity.

The next step is to calculate $S_i(0)$ for each set of ${\bf x}_i(0)$, ordering them according to increasing values of $S_i$. Then the  first $N_\rmn{elite}$ set of parameters is selected, i.e. the $N_\rmn{elite}$-samples with lowest $S$ values, which will be labelled as the elite sample array ${\bf X}^\rmn{elite}(0)$.

We then determine  the mean and standard deviation of the elite sample, $\bar{p}^\rmn{elite}_j(0)$ and ${\bf \sigma}^\rmn{elite}_j(0)$ respectively, as:

\begin{equation}
  \bar{p}^\rmn{elite}_j(0)=\frac{1}{N_\rmn{elite}}\sum\limits_{i=1}^{N_\rmn{elite}}X^\rmn{elite}_{ij}(0),
\end{equation}

\begin{equation}
  {\bf \sigma}^\rmn{elite}_j(0)=\sqrt{\frac{1}{\left(N_\rmn{elite}-1\right)}\sum\limits_{i=1}^{N_\rmn{elite}}\left[X^\rmn{elite}_{ij}(0)-\bar{p}^\rmn{elite}_j(0)\right]^2}.
\end{equation}

The array ${\bf X}$ at the next iteration is determined as:

\begin{equation}
  X_{ij}(1)=\bar{p}^\rmn{elite}_j(0)+{\bf \sigma}^\rmn{elite}_j(0) G_{ij},
\end{equation}

This process is repeated from equation (14), with $G_{ij}$ regenerated at each iteration. The optimization stops when the maximum number of iterations $k_\rmn{max}$ is reached.

In order to prevent convergence to a sub-optimal solution due to the intrinsic rapid convergence of the CE method, \citet{kro06} suggested the implementation of a fixed smoothing scheme for $\bar{p}^\rmn{elite,s}_j(k)$ and ${\bf \sigma}^\rmn{elite,s}_j(k)$:

\begin{equation}
  \bar{p}^\rmn{elite,s}_j(k)=\alpha^\prime\bar{p}^\rmn{elite}_j(k)+\left(1-\alpha^\prime\right)\bar{p}^\rmn{elite}_j(k-1),
\end{equation}

\begin{equation}
  {\bf \sigma}^\rmn{elite,s}_j(k)=\alpha_\rmn{d}(k){\bf \sigma}^\rmn{elite}_j(k)+\left[1-\alpha_\rmn{d}(k)\right]{\bf \sigma}^\rmn{elite}_j(k-1),
\end{equation}
where $\alpha^\prime$ is a smoothing constant parameter ($0<\alpha^\prime< 1$) and $\alpha_\rmn{d}(k)$ is a dynamic smoothing parameter at $k$th iteration:

\begin{equation}
  \alpha_\rmn{d}(k)=\alpha-\alpha\left(1-k^{-1}\right)^q,
\end{equation}
with $0<\alpha< 1$ and $q$ is an integer typically between 5 and 10 \citep{kro06}.

Following \citet{cap09}, we adopted $\alpha^\prime=1$, $\alpha=0.7$ and $q=5$ throughout this work. In addition, we also assumed $N=100$, $N_\rmn{elite}=10$ and $k_\rmn{max}=800$. It is important to emphasize that this value for $k_\rmn{max}$ was chosen so that, independently of the number of CE optimizations that are performed for a fixed $P_\rmn{prec,obs}$ and sense of precession, the remaining precession parameters will always converge to the same numerical value (variations not larger than a few per cent). For the present work, we run our code three times for each particular precession period and sense of precession.

\subsection{The merit function $S$}

The merit function $S(k)$ transmits to the CE algorithm the best tentative set of precession parameters at the $k$th iteration. Following \citet{cap09}, we have chosen $S(k)$ as:

\begin{equation}     
   S(k)= \Upsilon(k)+\sum\limits_{i=1}^{N_\rmn{d}}\left\{\left[S_{\alpha_i}(k)\right]^2+\left[S_{\delta_i}(k)\right]^2+\left[S_{r_i}(k)\right]^2\right\},
\end{equation} 
where:

\begin{equation}     
   S_{\alpha_i}(k)= \Delta\alpha_i-\Delta\alpha_\rmn{mod_i}(k),
\end{equation}

\begin{equation}     
   S_{\delta_i}(k)= \Delta\delta_i-\Delta\delta_\rmn{mod_i}(k),
\end{equation}

\begin{equation}     
   S_{r_i}(k)= \Delta r_i-\Delta r_\rmn{mod_i}(k),
\end{equation}
where $\Delta\alpha_i$ and $\Delta\delta_i$ are, respectively, the right ascension and declination offsets of the jet knot $i$, $\Delta r_i^2=\Delta\alpha_i^2+\Delta\delta_i^2$.

Our code generates, for each observation data $t_\rmn{obs}$, pairs of right ascension and declination offsets based on different values of $t_\rmn{ej}$, which are provided from a given jet-precession model. The chosen model (RA,Dec)-pair is that which minimizes the square distance between the precession helix and the observed (RA,Dec.) pair, also respecting the condition $t_\rmn{obs,ej}<t_\rmn{obs}$. Therefore, for each observed (RA,Dec.) pair there is always a $t_\rmn{obs,ej}$ that makes $S_{\alpha_i}^2+S_{\delta_i}^2$ a minimum.

As mentioned in \citet{cap09}, the terms $S_{\alpha_i}$ and $S_{\delta_i}$ strongly constrain the instantaneous jet direction in the optimization process, while inclusion of $S_{r_i}$ provides additional constraint on the modelling of the core-component distances as well as improving the convergence performance of the method. 

Jet kinematic modelling performed only on the plane of the sky has an extra potential difficulty: the jet components are not fully independent of each other, in the sense that unique identifications of jet knots also depend on estimates of their individual proper motions. However, identification problems related to fitting procedures, as well as observations poorly sampled in time, may influence the follow-up of the components in time, which consequently might contribute to a misinterpretation of the data. Our CE modelling avoids such potential biases, purely analysing the time behaviour of the sky position of the jet knots without any information concerning previous identification of components. To guarantee that our jet precession modelling will also predict apparent velocities similar to those observed in BL Lacertae, we included a penalty function $\Upsilon(k)$ in the CE optimization process \citep{cap09}. Based on independent estimates of the typical apparent velocities of the jet knots in BL Lacertae (e.g., \citealt{mut90,jor05,lis09}), we decided to include in the CE algorithm the following penalty function in equation (21):

\begin{equation}     
  \Upsilon(k) = \left\{\begin{array}{rl}
10,&\mbox{if}\quad \beta_\rmn{obs}^\rmn{min}(k)<2\quad \mbox{or}\quad \beta_\rmn{obs}^\rmn{max}(k)>11,\\
0,&\mbox{elsewhere},
\end{array}\right.
\end{equation}
where $\beta_\rmn{obs}^\rmn{min}(k)$ and $\beta_\rmn{obs}^\rmn{max}(k)$ are respectively the minimum and maximum values of the apparent jet speeds predicted by a given precession model at iteration $k$. It is important to emphasize that the choice of $\Upsilon(k) = 10$ mas$^2$ is sufficient to guarantee that any tentative solution providing $\beta_\rmn{obs}^\rmn{min}(k)<2$ or $\beta_\rmn{obs}^\rmn{max}(k)>11$ is statistically non-favoured during optimization. 

Note that the choice of $S$ is based on the minimization of the quadratic distances between the observational data and those produced by the precession model, while $\Upsilon(k)$ provides extra constraints on the parameters $\gamma$ and $\phi_0$.

%-----------------------------------------------------------FIGURE 01 
   \begin{figure}
   \begin{center}
	  {\includegraphics[width=60mm]{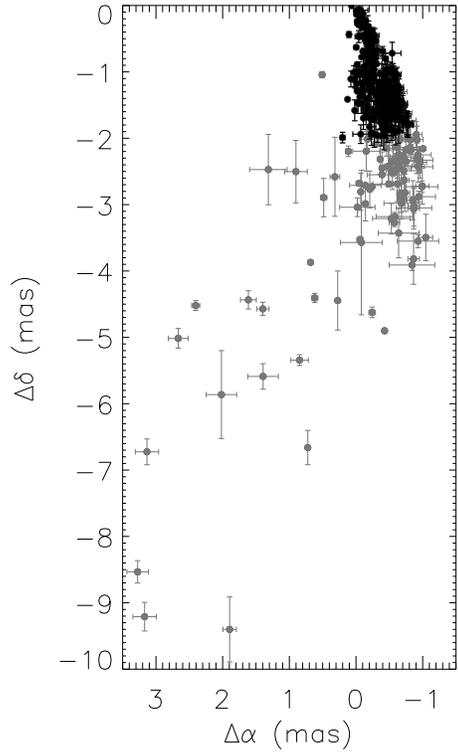}}
      \caption{The distribution of right ascension and declination offsets of the jet components of BL Lacertae. Grey circles represent the whole data extracted from the literature, while the black circles are the data analysed by our precession model (core-component distances smaller than 2 mas).}
      \label{BLLacdata}
	  \end{center}
   \end{figure}
%______________________________________________________________

%-----------------------------------------------------------FIGURE 02 
   \begin{figure*}
	  {\includegraphics[width=178mm]{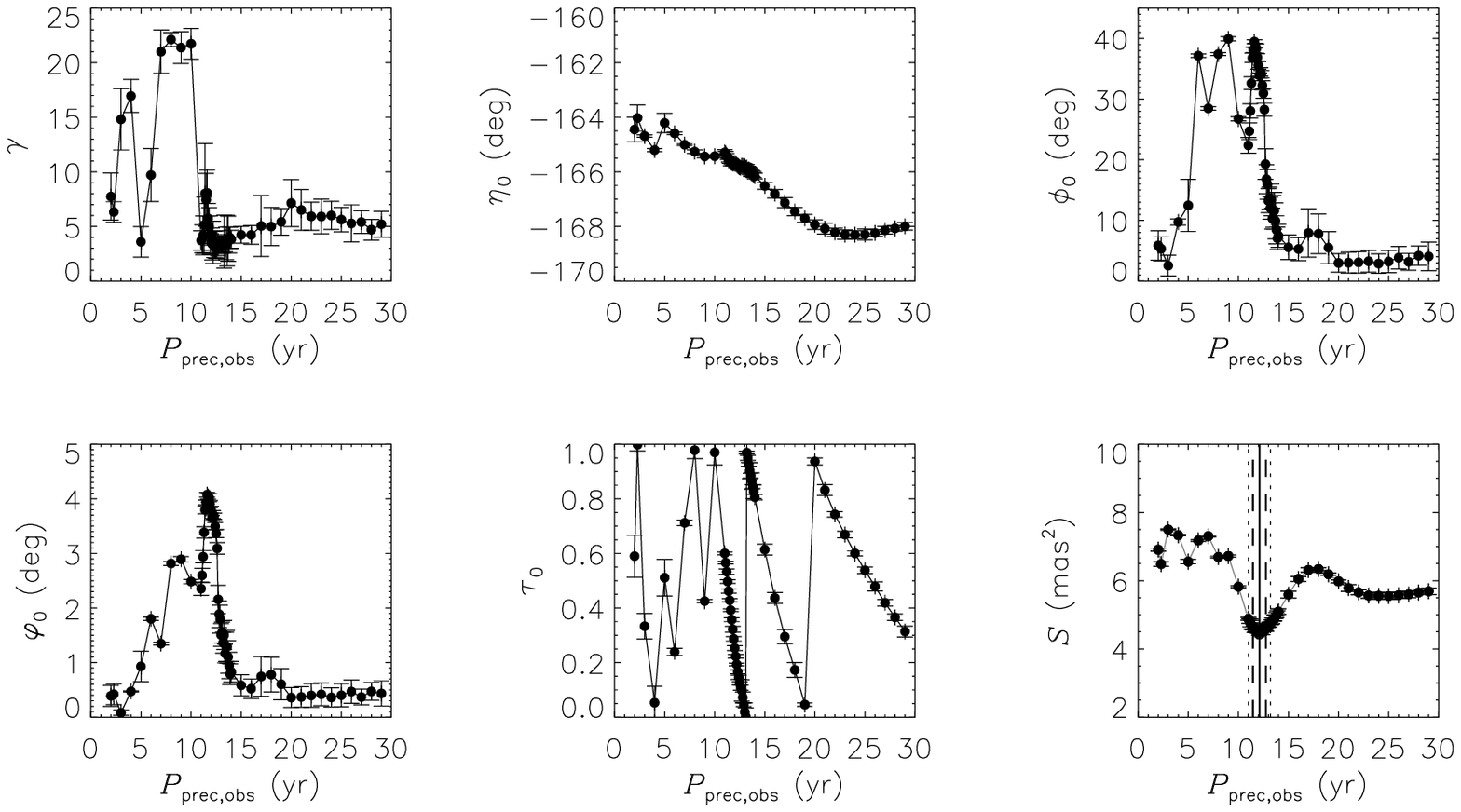}}
      \caption{Results for clockwise sense of precession ($\iota=1$). Dependence of the CE optimized precession model parameters in terms of the observed precession period, as well as the value of the merit function (lower rightmost panel). Steps of 1 yr in precession period were considered, except for the interval between 11 and 14 years, during which differences between consecutive precession periods of 0.1 yr were used. We also included in our analyses the value of 2.29 yr obtained by \citet{sti03}. The merit function reaches its minimum at a precession period between 11 and 13.5 yr (dotted vertical lines). The solid and dotted vertical lines mark respectively the weighted mean and standard deviation of the precession period related to the merit-function's minimum (12.11 $\pm$ 0.65 yr).}
      \label{precmodelparamcw}
   \end{figure*}
%______________________________________________________________

%-----------------------------------------------------------FIGURE 03 
   \begin{figure*}
	  {\includegraphics[width=178mm]{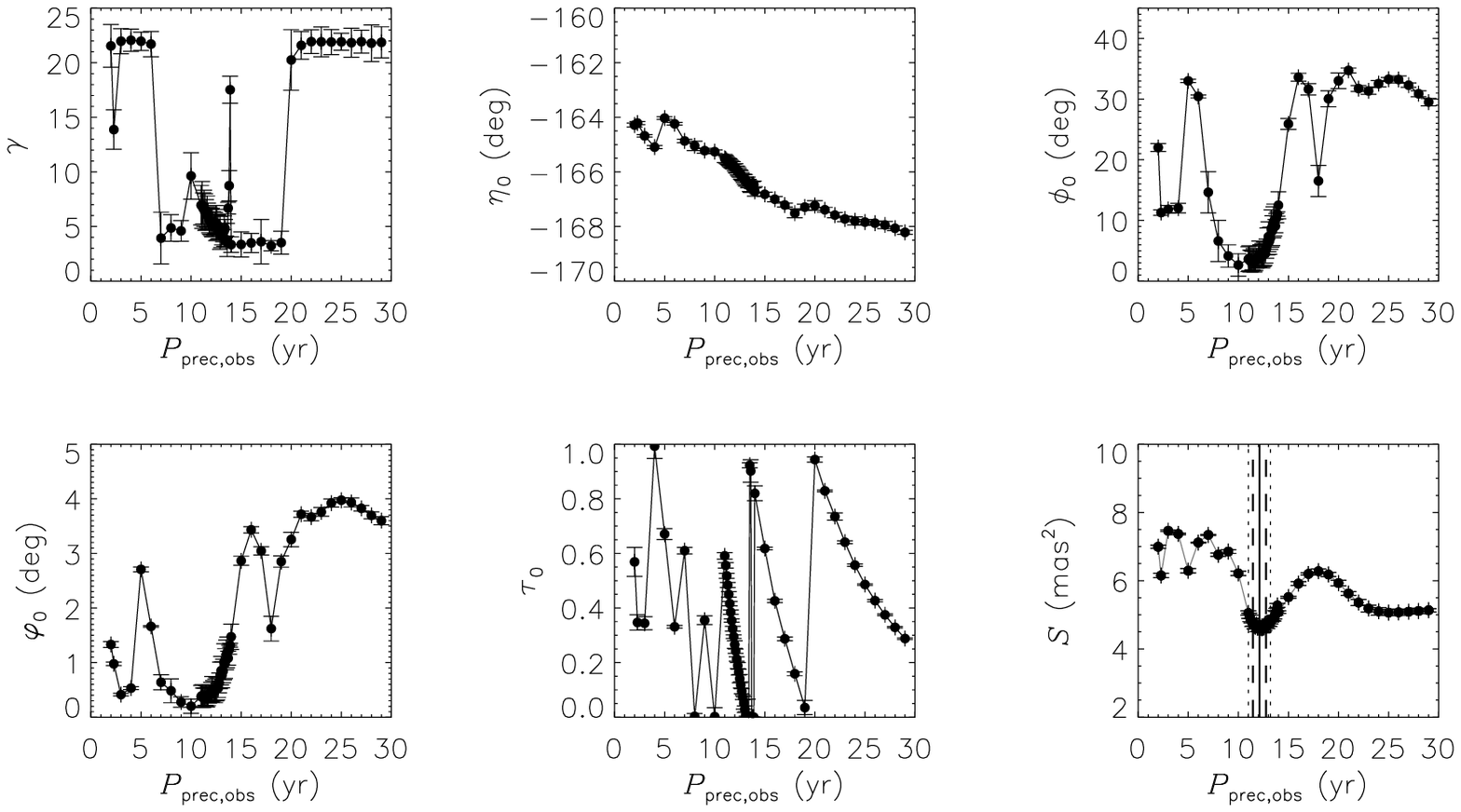}}
      \caption{Results for counterclockwise sense of precession ($\iota=-1$). Dependence of the CE optimized precession model parameters in terms of the observed precession period, as well as the value of the merit function (lower rightmost panel). Steps of 1 yr in precession period were considered, except for the interval between 11 and 14 years, during which differences between consecutive precession periods of 0.1 yr were used. We also included in our analyses the value of 2.29 yr obtained by \citet{sti03}. The merit function reaches its minimum at a precession period between 11 and 13.5 yr (dotted vertical lines). The solid and dotted vertical lines mark respectively the weighted mean and standard deviation of the precession period related to the merit-function's minimum (12.11 $\pm$ 0.65 yr).}
      \label{precmodelparamacw}
   \end{figure*}
%______________________________________________________________

\section{The jet precession model for the parsec-scale jet of BL Lacertae}

\subsection{Observational data}

The pc-scale jet observational data of BL Lacertae analysed in this work were gathered from the literature \citep{pere88,char90,mut90,bon96,gaca96,tat98,den00,fom00,gaca03,sti03,lis09}. These data (RA and Dec. offsets) were obtained at frequencies of 5.0, 8.4, 10.7 15, 24 and 43 GHz, from 1980.93 to 2007.682, providing a time monitoring of the jet activity in BL Lacertae of about 26.7 yr. 

We show in Fig. \ref{BLLacdata} the corresponding spatial distribution of the jet components of BL Lacertae on the plane of the sky. As already noted in previous papers (e.g., \citealt{char90,den00,sti03}), the pc-scale jet of BL Lacertae bends systematically to south-east after a core distance of about 3 or 4 mas, which definitely cannot be addressed exclusively by a jet precession phenomenon. Because of this, we decided to restrict our precession analysis to distances below of 2 mas (see filled circles in Fig. \ref{BLLacdata}), avoiding the bending zone of the jet.

Given the multiwavelength nature of the data used in this work, opacity effects are expected to affect the determination of the absolute core position, which introduce frequency-dependent shifts in the core-component distances and proper motions (e.g., \citealt{blko79,loba98,kov08}). In the case of a precessing jet, these corrections are time-dependent since the angle between the jet and the line-of-sight varies with time \citep{caab04a,caab04b}. In the case of BL Lacertae, \citet{mut90} found 0.3 mas for the magnitude of the core-shift between 5.0 and 10.6 GHz. \citet{den00} estimated a core-component shift between 22 and 43 GHz smaller than 0.1 mas. \citet{crga08} calculated a core shift of 0.20 mas comparing 5.1- and 7.9-GHz maps, while \citet{osga09}, using images at frequencies between 4.6 to 43.1 GHz, found core-shifts values below of 0.43 mas. Considering these works, the mean value of the core-shift in BL Lacertae is roughly 0.2 mas, which will not influence significantly our jet precession modelling given the intrinsic uncertainties of the data, as well as the high fraction of 15-GHz data in our sample\footnote{We run additional CE optimizations considering two subsets of the original data to verify quantitatively this claim. The two subsets correspond to 15-GHz and 15+43 GHz data (about 44 and 83 per cent of the original data set, respectively). For 15-GHz data, typical changes in the values of the precession parameters in relation to our best model (see Table 1 in Section 4.4) are usually less than $\sim$7 per cent (not exceeding $\sim$25 per cent). For 15+43 GHz data, the typical differences are less than $\sim$3 per cent ($\sim$15 per cent in the worst case). The decrease in those values is a consequence of the increase of the time coverage of 15+43 GHz data (from $\sim$ 8.3 yr to $\sim$ 13 yr).}. Because of that, we will hereafter neglect core opacity effect in our jet precession analyses.

%-----------------------------------------------------------FIGURE 04 
   \begin{figure}
   \begin{center}
	  {\includegraphics[width=63mm]{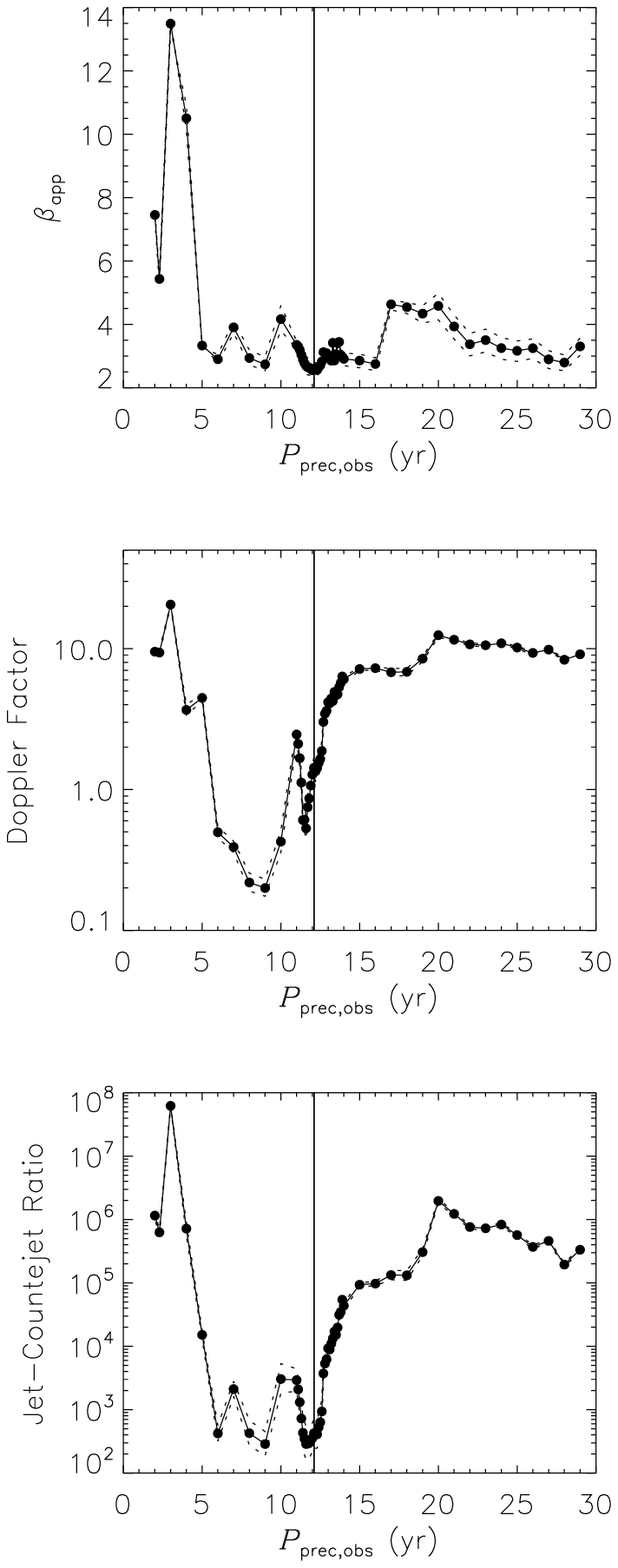}}
      \caption{Results for clockwise sense of precession ($\iota=1$). Apparent velocity, Doppler-boosting factor and jet-counterjet ratio (assuming $p=2$ and $\alpha=0.8$) predicted by our CE optimized model as a function of the precession period. Full circles represent those quantities calculated at $\phi=\phi_0$, while the dotted lines show their respective upper and lower limits allowed by the precession model. The solid vertical line marks the favoured precession period of 12.1 yr.}
      \label{precperiodcw}
	  \end{center}
   \end{figure}
%______________________________________________________________

%-----------------------------------------------------------FIGURE 05 
   \begin{figure}
   \begin{center}
	  {\includegraphics[width=63mm]{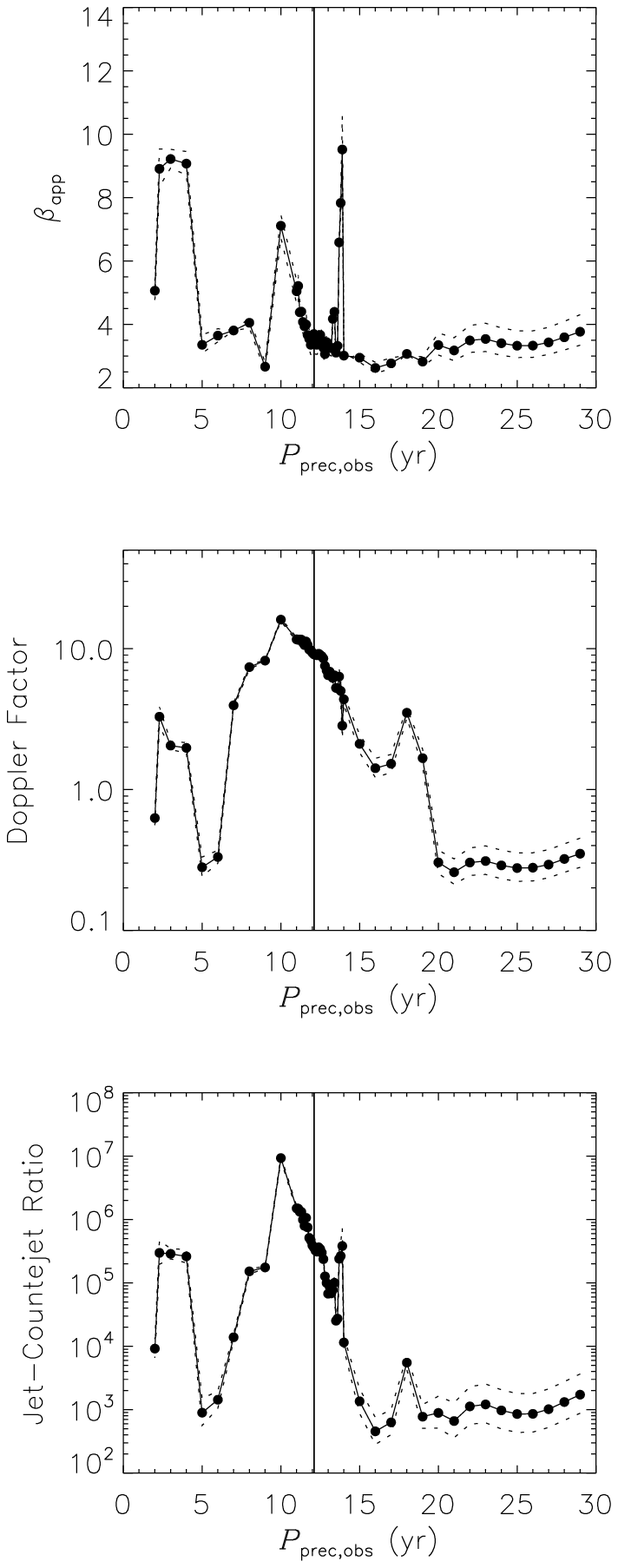}}
      \caption{Results for counterclockwise sense of precession ($\iota=-1$). Apparent velocity, Doppler-boosting factor and jet-counterjet ratio (assuming $p=2$ and $\alpha=0.8$) predicted by our CE optimized model as a function of the precession period. Full circles represent those quantities calculated at $\phi=\phi_0$, while the dotted lines show their respective upper and lower limits allowed by the precession model. The solid vertical line marks the favoured precession period of 12.1 yr.}
      \label{precperiodacw}
	  \end{center}
   \end{figure}
%______________________________________________________________

\subsection{Estimating the precession period and the sense of precession of the BL Lacertae jet}

The precession period and the sense of precession are quantities not automatically optimized by our CE method. Thus, some extra procedure is necessary to determine those parameters. Following \citet{cap09}, we mapped the dependence of the jet-precession model fitting with the precession period and the sense of precession. We varied the value of the precession period measured in the observer's reference frame from 2 to 30 yr in steps of 1 yr for both clockwise and counterclockwise senses of precession, covering those values previously suggested in the literature \citep{sti03,tate09}. An extra refinement step of 0.1 yr was adopted between 11 and 14 yr, in which the merit function presented a minimum, as will be discussed later.

For each tentative precession period, our CE method searched for the best set of precession model parameters in the ranges: $0.95\leq \beta \leq 0.999$ ($3.2\leq \gamma \leq 22.4$), $-180\degr \leq \eta_0 \leq -150\degr$, $0\fdg 1\leq \phi_0\leq 40\degr$, $0\degr \leq \varphi_0 \leq 30\degr$ and $0\leq \tau_\rmn{0,s} \leq 1$. It is important to emphasize that these parameter ranges contain the jet parameter values for BL Lacertae derived in previous works (e.g., \citealt{mut90,den00,sti03,tate09}). Note also that the hypothesis of a non-precessing jet is incorporated in our analysis, since $\varphi_0=0\degr$ is one of the possibilities to fit the observational data.

We show in Figs. \ref{precmodelparamcw} and \ref{precmodelparamacw} the optimized precession model parameters and the merit function versus the precession period in the observer's reference frame, for clockwise and counterclockwise senses of precession, respectively. Among all precession model parameters, $\eta_0$ exhibits the smallest variation in terms of $P_\rmn{prec,obs}$ (only about 4$\degr$). Extreme values for $\gamma$ and $\phi_0$ were found for some tentative precession periods; they correspond to $ 5<P_\rmn{prec,obs}<14 $ for clockwise precession, and $P_\rmn{prec,obs}<5$ or $ P_\rmn{prec,obs} >19 $ for counterclockwise precession.

A well-defined minimum for the merit function is found at a precession period of about 12.1 years, independently of the sense of precession. 
The values of both minima are practically identical ($S=4.511\pm 0.109$ mas$^2$ and $S=4.606\pm 0.072$ mas$^2$ for clockwise and 
counterclockwise precession, respectively), making impossible to obtain information about the sense of precession from the merit function alone. The difficulty in detecting the correct sense of precession was also found by \citet{sti03}. As mentioned by them, the problem probably resides on the lack of observational data with larger core-component distances employed in the modelling. However, other constrains on the optimized precession parameters can be used instead, namely the interval of observed superluminal velocities, the boosting parameter, which depends on the Doppler factor, and the jet-to-counterjet flux density ratio $\Xi$, which should be large enough to guarantee that the counterjet is not observed. This last quantity can be calculated from:

\begin{equation}   
\Xi(\tau_\rmn{s})=\left[\frac{1+\beta\cos\phi(\tau_\rmn{s})}{1-\beta\cos\phi(\tau_\rmn{s})}\right]^{p+\alpha},
\end{equation}
where $p$ is equal to 2 or 3 for a continuous or clumpy jet, respectively (e.g., \citealt{libl85}), and $\alpha$ is the spectral index of the flux density distribution in terms of the observed frequency $\nu$ ($S_\nu \propto \nu^{-\alpha}$).

We plotted the dependency of the interval of predicted superluminal velocities, Doppler  factor and jet-counterjet ratio as a function of the precession period in the observer's reference frame in Figs. \ref{precperiodcw} (clockwise sense of precession) and \ref{precperiodacw} (counterclockwise sense). Looking carefully at $P_\mathrm{prec,obs}\sim 12.1$ yr we can see that the clockwise jet-precession model predicts too low value of $\Xi$ ($\sim 300$), which implies the possibility of counterjet detection in interferometric radio images, given their typical dynamic  ranges. However, such detection has never been achieved, which indicates that a clockwise jet-precession scenario is not appropriated for BL Lacertae. On the other hand, the counterclockwise jet-precession model predicts $\Xi\sim 3.5\times 10^5$,  in agreement with available observational data. 

In addition, our counterclockwise precession model predicts Doppler-boosting factor ranging from 8.8 and 9.4 (in contrast to the clockwise model that predicts a variation between 1.2 and 1.7), in good agreement with independent estimates of the Doppler-boosting parameter for BL Lacertae found in the literature that suggest $\delta \ga 7$ (e.g., \citealt{jor05,hov09,wu11}). Therefore, the counterclockwise 12.1-yr jet precession scenario is favoured in the case of BL Lacertae. 

The 12.1-precession period found in this present work is not in agreement with the precession periods of 2.29 and 26.0 years reported previously by \citet{sti03} and \citet{tate09}, respectively. However, \citet{mude05}, analysing a different set of 43-GHz images (including some overlapping epochs with \citealt{sti03}), concluded that changes in the structural position angle\footnote{Angle defined by the relative orientation between components C1 and C2 detected in the 43-GHz VLBI images of BL Lacertae (see \citealt{sti03}).} of the radio core of BL Lacertae could present a periodicity of 12.1 yr, as statistically significant as that of 2.29 yr. Note that all these previous works analysed a substantially shorter interval of observations ($\sim$3.1 years in \citealt{sti03}, $\sim$5.1 years in \citealt{mude05} and $\sim$ 12 yr in \citealt{tate09}) in comparison to this present work ($\sim$26.7 yr), which might be responsible for such discrepancies. Another possibility is that the 2.29-yr period is related to some extra phenomenon, such as nodding modulation of the precession angle, for instance (see Section 6 for a more detailed discussion about this possibility).

\subsection{The non-precessing jet scenario}

As mentioned before, the non-precessing jet hypothesis ($\varphi_0=0\degr$) was also checked by our CE technique during each optimization processes. Even though the validation  tests \citep{cap09} showed that the technique is able to determine whether precession is taking place, we decided  to perform an independent model optimization, forcing $\varphi_0=0\degr$, and rerun the code to optimize the remaining jet model parameters. We found that the value of merit function (defined by equation 21) in this situation corresponds to $\sim$7.8 mas$^2$, at least 1.5 times larger than that found for $P_\mathrm{prec,obs}\sim 12.1$ yr. To estimate the statistical significance of such difference, we applied the non-parametric two-sample two-dimensional Kolmogorov$-$Smirnov test,  (2DKS test hereafter: see e.g., \citealt{peac83,fafr87,pre97}).

In brief, this test compares two-dimensional data sets in terms of the maximum cumulative difference between them, measuring the probability of these two data sets being drawn from the same parent distribution. In this work, the data sets are the observed and model predicted jet-knot position offsets from the core, so that comparison is made using pairs of ($\Delta\alpha_\rmn{obs_i}$, $\Delta\delta_\rmn{obs_i}$) and ($\Delta\alpha_\rmn{mod_i}$, $\Delta\delta_\rmn{mod_i}$), the $i$th data point of the respective data sets. Note that these two samples are not strictly independent since the model coordinates are calculated from a model that was constrained, on the other hand, from the observed positions of the jet knots. Although it is in contradiction to the underlying hypothesis of independence of the two analysed data sets, \citet{wil05} argue if the number of the observed data points is much greater than the number of the model parameters, such effect should be small enough to allow the application of the 2DKS test. This requirement is fulfilled in our data sets. 

Unfortunately, another potential drawback concerning the multidimensional generalization of the Kolmogorov-Smirnov test is related to the calculation of the maximum cumulative difference in two or more dimensions, which is not uniquely defined (e.g., \citealt{peac83,fafr87,pre97}). It introduces a possible dependency in the determination of the maximum cumulative difference with the shape of the parent two-dimensional probability distribution. However, \citet{fafr87} showed from Monte Carlo simulations that their parametrisation for 2DKS test is distribution-free in good approximation. In any case, the resulting probabilities obtained from 2DKS tests are not as rigorously reliable as in the one-dimensional case, so that they must be interpreted warily (e.g., \citealt{pre97}).

The application of the 2DKS test for our counterclockwise 12.1-yr jet precession model leads to a probability of $\sim$15 per cent. As suggested in \citet{pre97} and \citet{wil05}, a 2DKS probability above  $\sim$10-20 per cent can be used to claim reasonable agreement between the two analysed data sets. Conversely, this probability drops to 0.028 per cent for the non-precessing jet scenario. Thus, the 2DKS tests indicate that the probability of the observed and model data sets being drawn from the same parent distribution is 15 per cent for our best precession model, but less than 0.1 per cent for the non-precession scenario. This demonstrates that our model provides a significantly better description of the collected observations for BL Lacertae than a non-precessing model, although some discrepancies between our best model and observational data still remain.

%-----------------------------------------------------------TABLE 01 

\begin{table}
 \centering
 %%\begin{minipage}{127mm}
  \caption{The best precession model parameters optimized by our CE algorithm for a counterclockwise sense of precession ($\iota=-1$). The uncertainties in each parameter correspond to $1\sigma$ level. \label{prec_mod_param}}
  \begin{tabular}{@{}lcccr@{}}
  \hline
$P_\rmn{prec,obs}$ (yr) & & & & 12.11 $\pm$ 0.65\\
$P_\rmn{prec,s}$  (yr)  & & & & 550 $\pm$ 247\\
$\beta$ & & & & 0.9824	$\pm$	0.0087\\
$\gamma$ & & & & 5.35	$\pm$	1.31\\
$\eta_0$ (deg) & & & & -165.86	$\pm$	0.16\\
$\phi_0$ (deg) & & & & 4.43	$\pm$	2.16\\
$\varphi_0$ (deg) & & & & 0.51	$\pm$	0.24\\
$\tau_\rmn{0,s}$ & & & & 0.242	$\pm$	0.008\\
\hline
\end{tabular}
%%\end{minipage}
\end{table}

%______________________________________________________________

\subsection{Model parameters for 12.1-years counterclockwise precession period}

As presented in last section, the precession of the pc-scale jet of BL Lacertae occurs in a counterclockwise sense, respecting a periodicity of about 12.1 yr in the observer's reference frame. The CE optimized precession model parameters related to this period are given in Table \ref{prec_mod_param}. Precession-model parameters and their respective uncertainties were estimated following \citet{cap11} (equations 10 and 11 in their paper). Note that the precession rate in the source's reference frame is substantially slower, as expected from relativistic effects (see equation 11).

Comparison between snapshots of the precession helix generated from the parameters listed in Table \ref{prec_mod_param} and the right ascension and declination offsets of the observed jet knots of BL Lacertae can be visualized in the section entitled Supporting Information, published in the online version of this work. The distribution of the residuals in right ascension and declination directions (generated from the difference between the model-predicted and observed positions of jet knots) is compatible with zero-mean value in both directions. Their associated root mean square values (rms) are $\sim 0.12$ and $\sim 0.03$ mas in right ascension and declination directions, respectively. It strongly indicates that the mean jet position angle predicted from our best model is quite well determined.  

Applying the same analysis for a non-precessing jet, it is possible to note an increase of the spread of the residuals in relation to that found from our best precession model. It is quantitatively corroborated by the slight increase of the rms in both directions ($\sim 0.15$ mas in right ascension and $\sim 0.04$ mas in declination), reinforcing our conclusion that the non-precessing jet scenario is not suitable for BL Lacertae. In relation to the previously published precession models by \citet{sti03} and \citet{tate09}, the amplitude of residuals is even higher in comparison with a non-precessing jet scenario and our best precession model. Indeed, this was already expected given their larger values of $S$ ($\sim 17$ mas$^2$ for \citealt{sti03} and 15 mas$^2$ for \citealt{tate09}, much higher than our own value for no-precession case).\footnote{In order to compare the merit function of our best jet-precession model quantitatively with those derived by \citet{sti03} and \citet{tate09},we generated precession helices based on the values of the parameters determined by those authors and compared them with the same observational data used in our work. Thus, these comparisons respect the same conditions used in this work (e.g. penalty function, jet region, frequency range, etc.), as well as their related small 2DKS test probabilities (0.039 per cent for \citealt{sti03} and $ 1.6\times 10^{-4}$ per cent for \citealt{tate09}).}

According to our precession model, the expected apparent velocities of the jet of BL Lacertae are always superluminal, ranging from about 3.4 to 4.0$c$. Components having such velocities have been observed in previous works (e.g., \citealt{mut90,jor05,gaca03,lis09}), even though some quasi-stationary components ($\beta_\rmn{obs}<<1$), as well as faster knots have been also detected in BL Lacertae (e.g., \citealt{jor05,lis09}). It should not be interpreted as a serious problem of our model since some jet components of BL Lacertae exhibit non-ballistic trajectories, which is not taken into account in our present parametrization. Indeed, the theory of linear perturbations, as well as numerical simulations, has shown that jet precession can induce instabilities in the jet flow \citep{hard00,hard01,hard02}. The observational counterparts of these jet instabilities may be seen as non-ballistic knots in VLBI images. Therefore, the scenario adopted in this work (ballistic motions plus jet precession) does not exclude the existence of helical motions in the jet of BL Lacertae. Note also that our predicted apparent velocities are different from those found by \citet{sti03} ($\beta_\rmn{obs}\sim 6.7$) and \citet{tate09} ($7< \beta_\rmn{obs}< 10$).

Because of precession, jet viewing angle varies approximately between 4$\degr$ and 5$\degr$, which is in reasonable agreement with the values derived from the characteristics of the variability in the multiwavelength light curves of BL Lacertae \citep{hov09,rai10,sav10,wu11}. The predicted jet viewing angles from our precession model are between those found by \citet{sti03} ($6\degr <\phi< 12\degr$) and \citet{tate09} ($1.5\degr <\phi< 3\degr$).

The jet inlet position angle on the plane of the sky ranges from -172$\degr$ to -159$\degr$, providing a total variation of $\sim 13\degr$ which is approximately half of those obtained by \citet{sti03} and \citet{tate09}. This angular amplitude means an amplitude variation of 0.46 mas at 2 mas from the core, which is about twelve and four times greater than typical observational uncertainties in the right ascension and declination positions of the jet components, respectively.

As mentioned previously, the Doppler-boosting parameter varies roughly from 8.8 to 9.4 in our model, which means that the jet-counterjet flux ratio ranges from 3.3$\times 10^5$ to 3.9$\times 10^5$ (considering a continuous jet with a spectral index of 0.8; e.g., \citealt{libl85}), implying in a non-detected counterjet at pc scales. This result also supports the hypothesis claimed by \citet{sti03} that component C1 is not the manifestation of the counterjet.

The observational data used to constrain our optimized precession model covers 26.7 yr of monitoring, which means an interval of about 2.2 times longer than the estimated precession period of 12.1 yr. At this point it is important to emphasize that our CE model technique is able to deal properly with short time samplings, as can be seen in the validation tests published in \citet{cap09}. One of these tests employed synthetic data formed by no more than 50 sky positions (about one sixth of the data analysed in this work) and covered only 1.5 precession periods. Therefore, this strongly suggests that our results are not influenced by the relatively short time covering of the data in terms of the 12.1-yr precession period.

\section{The underlying jet of BL Lacertae}

The jet precession period of 12.1 yr has no counterpart detected in analyses of the historical light curves of BL Lacertae conducted up to now. Quasi-periodic variations in the radio light curves of BL Lacertae have been reported in the literature, ranging roughly from 0.7 to 8 yr depending on the frequency and time coverage of the observations \citep{kel03,vil04}. In optical
domains the situation is almost the same: there is some evidence of a periodicity of about 7-8 yr but this is still inconclusive (\citealt{hag97,vil04}). Similarly, \citet{fan98} reported a strong signature of a periodicity of 13.97 yr in the B-band light curve, as well as shorter periods of 0.66 and 0.88 yr, also found by \citet{web88}.

Although some works have shown the viability of jet precession in producing detectable signatures in the light curves of some sources (e.g., \citealt{caab04a,caab04b}), it should not be expected for all precessing objects. As mentioned by \citet{mude05}, the main contribution to the observed flux of BL Lacertae may be due to the extended jet, which is not varying its orientation and consequently may be masking any precession signature. Another
possibility that cannot be excluded is that the underlying jet is intrinsically weak, so that even Doppler-boosting effects modulated by precession are not sufficiently strong to make its contribution detectable.

We compared the historical 8.4-GHz and $B$ band\footnote{Based on data taken and assembled by the WEBT collaboration and stored in the WEBT archive at the Osservatorio Astronomico di Torino - INAF (http://www.oato.inaf.it/blazars/webt/).} light curves with upper limits for the observed flux density of the underlying jet\footnote{The term underlying jet refers to a continuous or quasi-continuous distribution of plasma elements in the jet in which shocks and/or plasmons (jet components) propagates (in a similar sense as given by \citealt{libl85}).} predicted from our best model. We assumed simplistically that the intrinsic flux density of the underlying jet $S_\rmn{s}$ did not vary along the observation time, following the relation $S_\rmn{obs}(t_\rmn{obs})=S_\rmn{s}[\delta(t_\rmn{obs})]^{2+\alpha}$ (similarly to \citealt{caab04b}). We assumed $\alpha\sim 0.0$ for the radio regime \citep{den00} and $\alpha=2.0$ for the optical band \citep{bro89}. To keep the underlying jet contribution always below the measured total flux densities, $S_\rmn{s}$ must be lower than 18 mJy and 0.30 $\mu$Jy at radio and optical frequencies, respectively.

The small variation in the value of the Doppler-boosting factor (between $\sim$ 8.8 and 9.4) implies a small variation in the amplitude of the observed flux density of the underlying jet (between $\sim$ 0.18 and 0.23 mJy in the $B$ band and from about 1.39 and 1.59 Jy at 8.4 GHz). We speculate that it might be the reason for the non-detection of such 12.1 yr periodicity from the previous statistical analyses of the light curves of BL Lacertae. Notwithstanding, the underlying jet contribution might be responsible for a plateau-like structure (with mean values of about 0.21 mJy in the $B$ band and 1.49 Jy at 8.4 GHz) underneath of the main flux variations of BL Lacertae, which are probably produced by the jet components themselves; this can be interpreted as either shocks moving down the jet (e.g., \citealt{blko79,hug85,hug89}) or plasmons ejected from the core (e.g., \citealt{pake66}).

\section{Jet precession and nodding motions in BL Lacertae}

\citet{sti03} analysed the structural position angle defined by the relative orientation between components C1 and C2, as well as the polarization angle in radio and optical observations, finding a periodic modulation of 2.29 yr in both quantities. They interpreted those results as due to the precession of the jet inlet region. We explore in this section an alternative interpretation for this short period in terms of a nutation motion produced in a supermassive binary black hole system.

\citet{kat82} established the relation between the characteristics of a precessing accretion disc in close binary systems and the frequency $\omega_\rmn{n}$ of short-term nodding motions:

\begin{equation}       
\omega_\rmn{n}=2(\omega_\rmn{orb}-\omega_\rmn{s}),
\end{equation}
where $\omega_\rmn{orb}$ is the orbital angular velocity of the secondary black hole. Note that $\omega_\rmn{s}$ should be negative since the induced precession is retrograde, in the sense of being contrary to the rotation of the accretion disk (eg., \citealt{kat82}).

Assuming that the nodding oscillation has a period of 2.29 yr in the observer's reference frame, we can use equation (27) to calculate the value of the orbital period of the secondary black hole $P_\mathrm{orb,obs}$ to produce the inferred 12.1-yr precession period. It implies $P_\mathrm{orb,obs}=7.4\pm 1.8$ yr or $P_\mathrm{orb,s}=335\pm 173$ yr in the source's reference frame, the respective uncertainties having been obtained from error propagation of the involved periods.

From the $P_\mathrm{orb,s}$, we can derive the distance between the two black holes $R_\mathrm{ps}$ from the Kepler's third law:

\begin{equation}
   R_\mathrm{ps}^3 = \frac{\mathrm{G}M_\rmn{tot}}{4\pi^2}P_\mathrm{orb,s}^2,
\end{equation}
where $G$ is the gravitational constant, and $M_\rmn{tot} = M_\mathrm{p}+M_\mathrm{s}$, the sum of the masses of the primary and secondary black holes, respectively. 

Several works have attempted to estimate the mass of the supermassive black hole in BL Lacertae \citep{fan99,wuur02,mahu03,ghi10,cap10}, which seems to have about $1.5-6\times 10^8$ M$_{\sun}$. Assuming that $M_\rmn{tot}$ corresponds to the mean value of those estimates, i.e $M_\rmn{tot} = 3.75\times 10^8$ M$_{\sun}$, we calculated the distance between the black holes from equation (28), providing  $R_\mathrm{ps} = 0.17 \pm 0.07$ pc, which corresponds to $\sim$ 0.13 mas at the distance of BL Lacertae.

To put a lower limit for the ratio between the masses of the secondary and primary black holes, $q=M_\rmn{s}/M_\rmn{p}$, it is necessary to verify in which conditions the orbital period of the secondary can produce the inferred accretion disc/jet precession rate. Assuming that jet precession of BL Lacertae is induced by torques in the primary accretion disc due to a non-coplanar secondary black hole (e.g., \citealt{katz97,pate95,larw97,rom00,caab04a,caab04b,cap06}), and taking $R_\mathrm{prec}\leq R_\mathrm{out}$ \citep{rom00}, where $R_\mathrm{prec}$ and $R_\mathrm{out}$ are respectively the outer radii of the precessing part of the disc and the disc itself, we can write \citep{cap06}:

\begin{equation}
  \left(\frac{|P_\mathrm{prec,s}|}{P_\mathrm{orb,s}}\right)\cos\varphi_0 \geq \frac{4}{3}\left(\frac{5-n}{7-2n}\right)\left[\frac{(1+q)^{1/3}}{0.88q^{2/3}f(q)}\right]^{3/2},
\end{equation}
where $n$ is the polytropic index of the gas (e.g., $n=3/2$ for a non-relativistic gas and $n=3$ for the relativistic case), and the function $f(q)$ has the form \citep{eggl83}:

\begin{equation}
   f(q) = \frac{0.49q^{2/3}}{0.6q^{2/3}+\ln(1+q^{1/3})}.
\end{equation}

The value of the left-hand of equation (29) is completely defined by our precession and nutation models, while the right-hand of the same equation depends only on the parameter $q$. Assuming $n=3/2$, equation (29) is satisfied only if  $q \ga 5.75$, which implies $ M_\mathrm{p} \la 5.5\times 10^7$ M$_{\sun}$ and $ M_\mathrm{s} \ga 3.2\times 10^8$ M$_{\sun}$. It is important to emphasize that there is no problem concerning secondary black hole to be more massive than primary one. The word primary in our context only means that the observed jet is generated from the primary black hole. It is not new at all, since some previous papers had already claimed systems with more massive secondary black hole than primary ones (e.g., \citealt{rom00,bri01}).

The time stability of such system can be verified by calculating the timescale $\tau_\rmn{GW}$ for losses due to gravitational radiation from the expression \citep{beg80,shte83}:

\begin{equation}
   \tau_\rmn{GW} = \frac{5c^5}{256}\frac{r_\rmn{ps}^4}{(GM_\rmn{tot})^3}\frac{(1+q)^2}{q},
\end{equation}
which is minimized when $q=1$.

For the supermassive binary black hole system considered in this work, we have $\tau_\rmn{GW}\geq 70$ Gyr, which is much longer than the age of the Universe. This indicates no significant changes in the orbit of the secondary and consequently to jet precession rate during the interval of the observations used in this work.

Concerning the intrinsic amplitude of the nutation motion, our calculation based on equation (4) in \citet{katz97} leads to a value of about 0\fdg10. Its inclusion into our jet-precession model generates an additional amplitude variation at the position angle of the jet on the plane of sky of about $\pm$2\fdg5, which may be detected in the millimetre data, in principle.

We can see that a putative supermassive binary black hole system with the physical characteristics described above can drive not only the jet-precession period found in this work, but also a nutation with a period of 2.29 yr, which might be responsible for variations in the structural position angle of the
pc-scale radio core, as well as in the polarization angles reported by \citet{sti03} and \citet{mude05}.

\section{Conclusions}

\citet{sti03} and \citet{tate09} have claimed the existence of precession motions in the pc-scale jet of BL Lacertae of 2.29 and 26 yr, respectively. In this work we revisited this issue, investigating temporal changes of the observed right ascension and declination offsets of the jet knots in terms of our relativistic jet-precession model. 

Our precession model is characterized by seven parameters: precession period, jet bulk velocity, aperture angle of the precession cone, the angle
between the cone axis and the line of sight, position angle of the cone axis on the plane of the sky, sense of precession, and precession angular phase. These parameters were optimized via heuristic cross-entropy (CE) method, comparing the projected precession helix with the two-dimensional position of the jet components on the plane of the sky, following \citet{cap09}. The search for the best precession model for BL Lacertae is performed considering parameter ranges that contain the jet parameter values derived in previous works (e.g., \citealt{mut90,den00,sti03,tate09}), as well as the case of a non-precessing jet ($\varphi_0=0\degr$).

Our optimized best model is compatible with a jet having a bulk velocity of 0.9824$c$, which is precessing in a counterclockwise sense with a period of about 12.1 yr at the observer's reference frame ($\sim$ 550 yr at the source's referential framework), and changing its orientation in relation to the line of sight between approximately 3\fdg9 and 4\fdg9. The position angle of the precession cone axis on the plane of the sky is about -166$\degr$. It is important to emphasize that these precession parameters lead to  $S\approx 4.6$ mas$^2$, which is significantly lower than the values obtained for both a non-precessing model and the previously published precession models. 

Searches for periodic variation in the historical light curves of BL Lacertae have not revealed any signature of flux variability occurring at a periodicity of 12.1 yr (e.g., \citealt{web88,fan98,hag97,kel03,vil04}). We speculate that the non-detection of the periodicity of 12.1 yr might be attributed to small variations in the amplitude of the Doppler-boosting factor predicted by our model, which implies small variations of the observed flux density associated to the underlying jet. Nevertheless, the underlying jet contribution might be responsible for a plateau-like structure underneath of the main flux variations of BL Lacertae.

Assuming that jet precession has its origin in a supermassive binary black hole system, we show that the 2.3-yr periodic variation in the structural position angle of the VLBI core of BL Lacertae reported by Stirling et al. is compatible with a nutation phenomenon if $ q\ga 5.75$. The orbital period of the secondary black hole at the source's reference frame is approximately 335 yr, which means a distance between primary and secondary black holes of about 0.17 pc.

Further monitoring of the pc-scale activity of BL Lacertae is necessary to confirm the validity of our precession model, as well as the supermassive binary black hole scenario proposed in this work.

\section*{Acknowledgments}

This work was supported by the Brazilian agencies Funda\c{c}\~ao de Amparo \`a Pesquisa do Estado de S\~ao Paulo (FAPESP) and Conselho Nacional de Desenvolvimento Cient\'\i fico e Tecnol\'ogico (CNPq). We thank Dr Massimo Villata for making available the WEBT $B$-band photometric data of BL Lacertae used in this work. This research has also made use of data, kindly provided by Dr Margo Aller from the University of Michigan Radio Astronomy Observatory, which has been supported by the University of Michigan and by a series of grants from the National Science Foundation, most recently AST-0607523, as well as NASA Fermi grants NNX09AU16G, NNX10AP16G, and NNX11AO13G. We also thank the anonymous referee for a detailed and careful report that improved substantially the presentation of this work.

\bsp

\label{lastpage}

\section{Supporting information}

Additional Supporting Information may be found in the online
version of this article:

This additional material shows the projections on the plane of the sky of the precession helices generated from the model parameters listed in Table \ref{prec_mod_param} over the entire observational period used in this work. We also display the 311 sky position of the jet knots of BL Lacertae that constrained our precession model. Note that precession helices describe the general trend of the observational data appropriately.

%-----------------------------------------------------------FIGURE 10 
   \begin{figure*}
	  {\includegraphics[width=150mm]{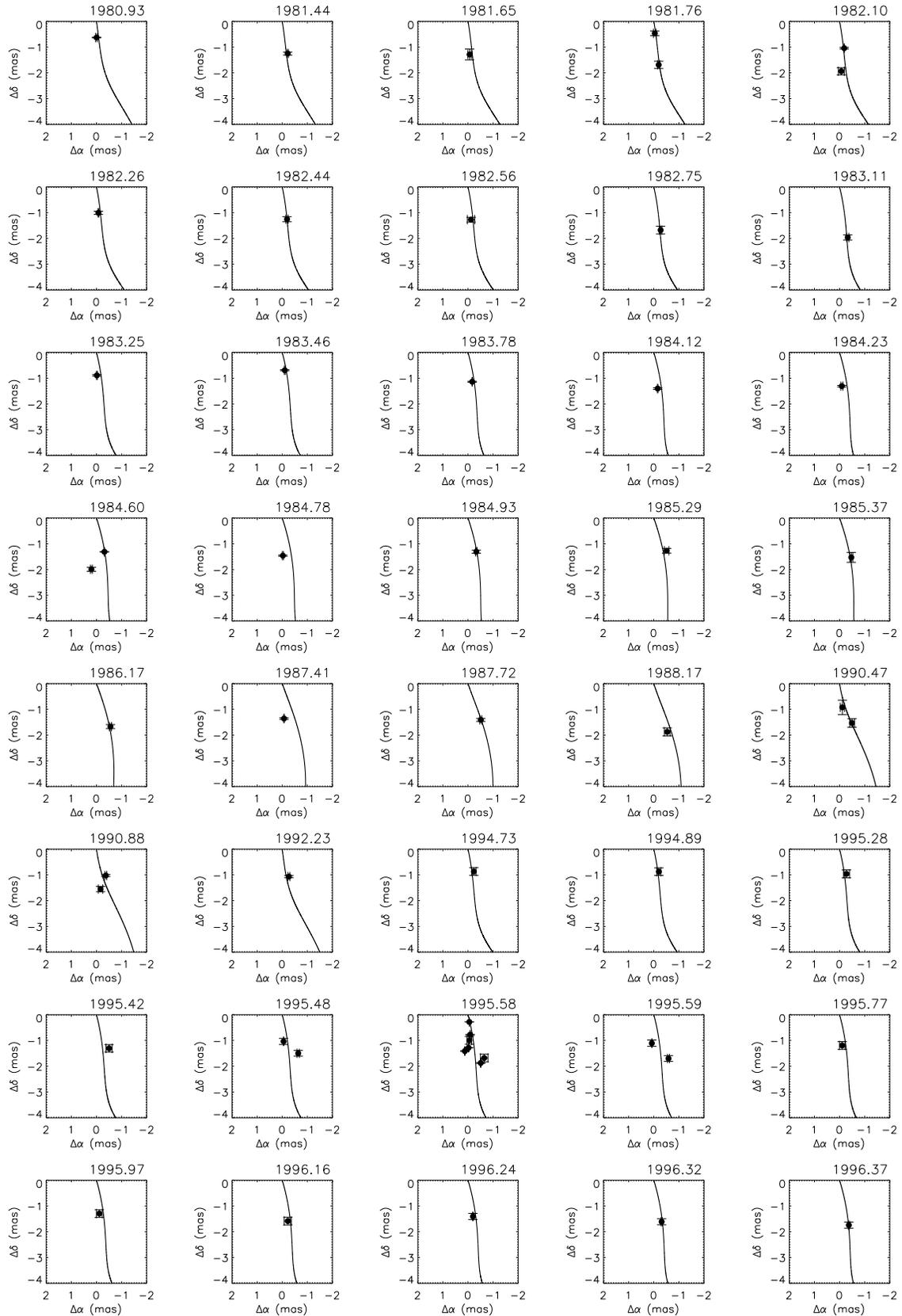}}
      \caption{Changes in the jet orientation of BL Lacertae along the time due to the precession model parameters listed in Table \ref{prec_mod_param}. Solid lines represent snapshots of the precession helix, while the positions of the jet knots and their respective uncertainties are shown by black dots. Numbers at upper right corners of the panels are the corresponding observation epochs.}
      \label{prec_helix}
   \end{figure*}

\setcounter{figure}{5}
   \begin{figure*}
	  {\includegraphics[width=150mm]{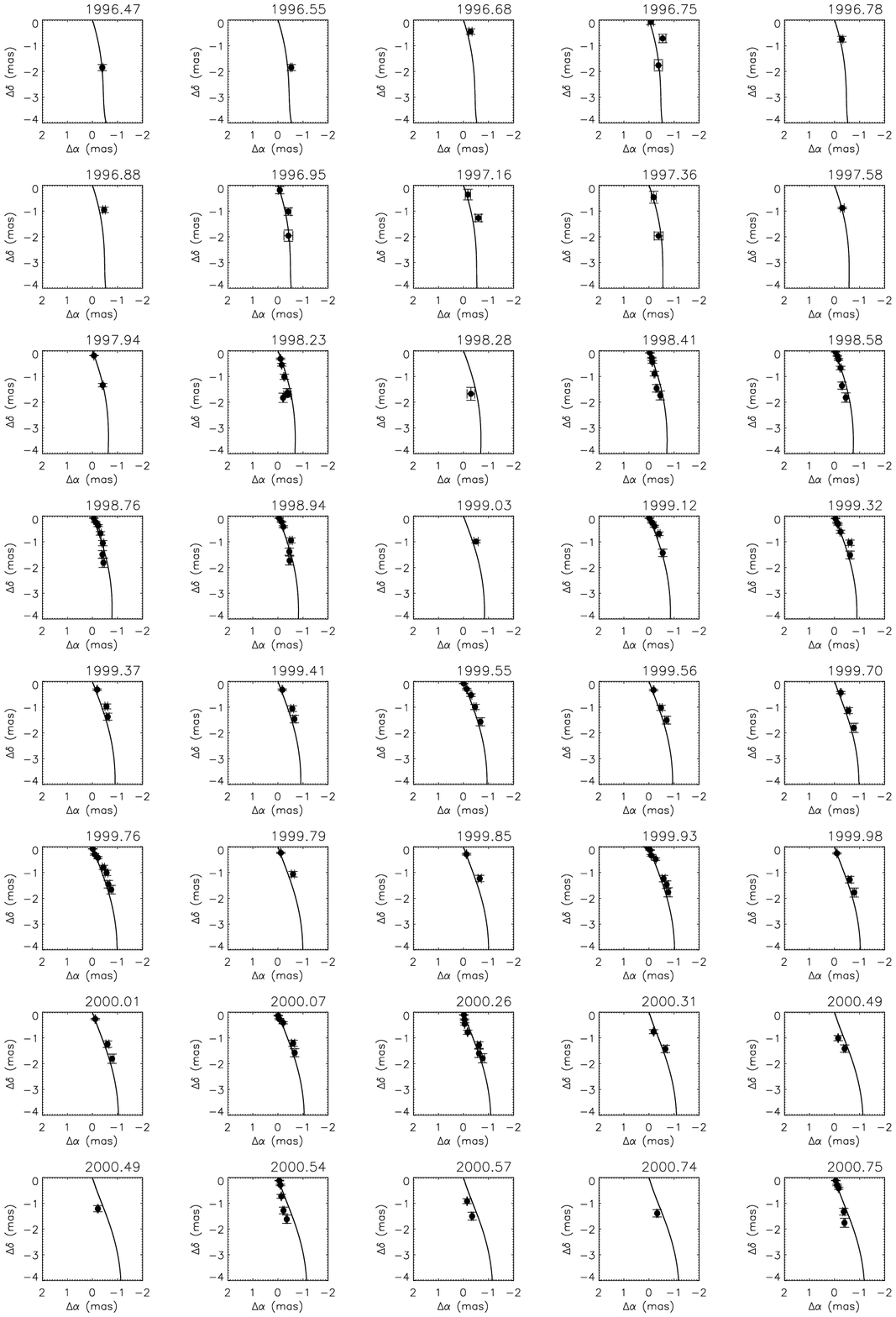}}
      \caption{Continued.}
      %\label{prec_helix}
   \end{figure*}

\setcounter{figure}{5}
   \begin{figure*}
	  {\includegraphics[width=150mm]{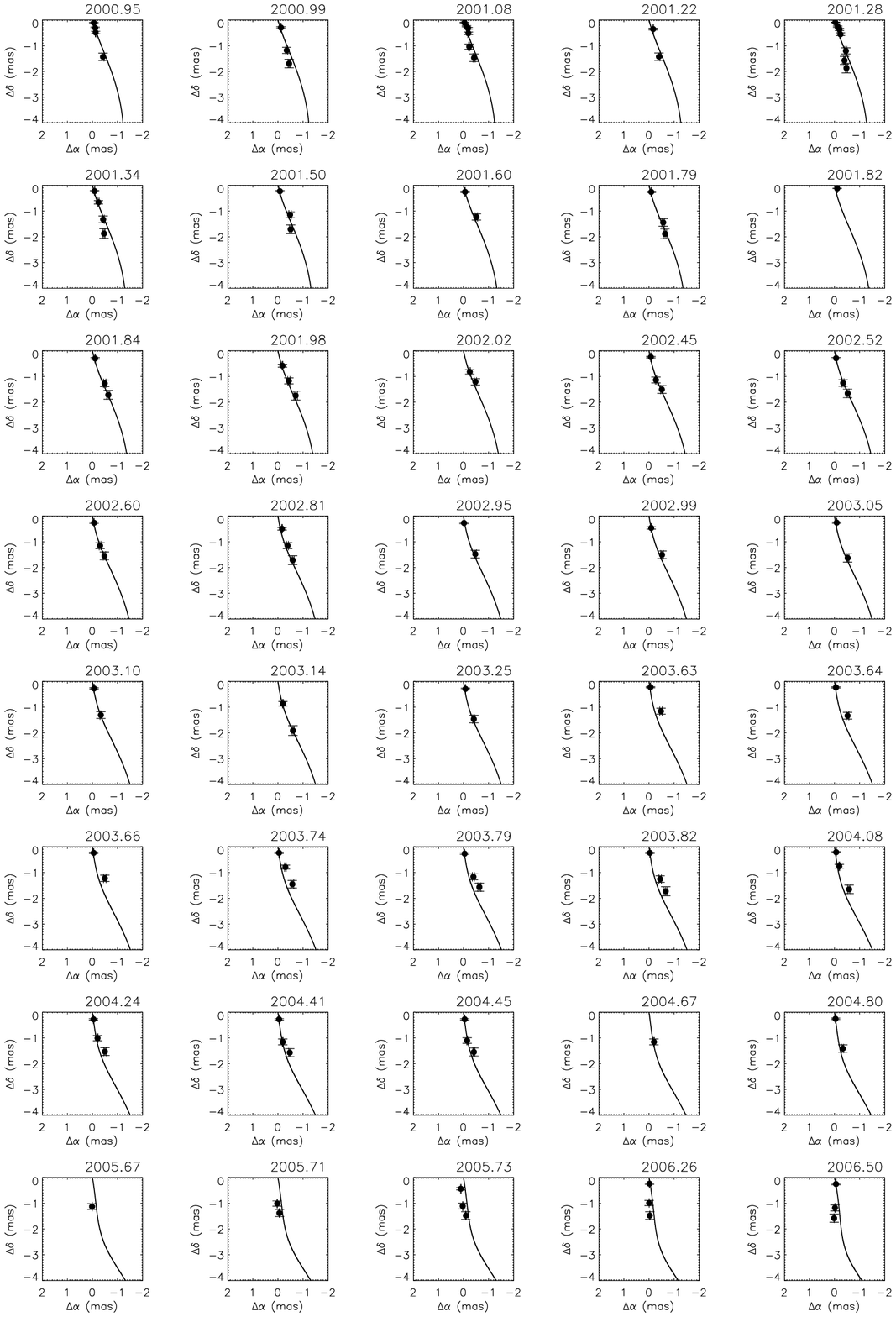}}
      \caption{Continued.}
      %\label{prec_helix}
   \end{figure*}

\setcounter{figure}{5}
   \begin{figure*}
	  {\includegraphics[width=150mm]{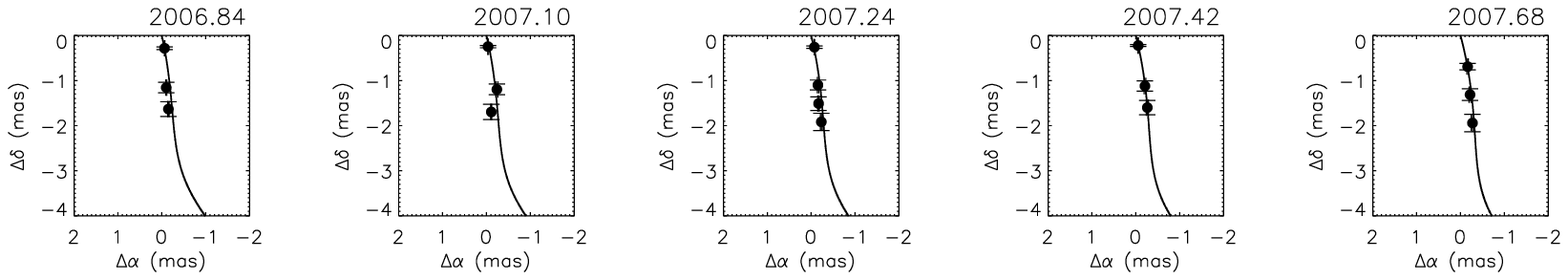}}
      \caption{Continued.}
      %\label{prec_helix}
   \end{figure*}

%______________________________________________________________


\begin{thebibliography}{99}

\bibitem[\protect\citeauthoryear{Antonucci}{1986}]{anto86} Antonucci, R. R. J. 1986, 
      ApJ, 304, 634

\bibitem[\protect\citeauthoryear{Begelman, Blandford \& Rees}{1980}]{beg80} Begelman, M. C., Blandford, R. D., Rees, M. J. 1980, 
      Nature, 287, 307.

\bibitem[\protect\citeauthoryear{Blandford \& K\"onigl}{1979}]{blko79} Blandford, R. D., K\"onigl, A. 1979, 
      ApJ, 232, 34
	  
\bibitem[\protect\citeauthoryear{Bondi et al.}{1996}]{bon96} Bondi, M., Padrielli, L., Fanti, R., Ficarra, A., Gregorini, L., Mantovani, F., Bartel, N., Romney, J. D., et al. 1996, 
      A\&A, 308, 415

\bibitem[\protect\citeauthoryear{Britzen et al.}{2001}]{bri01} Britzen, S., Roland, J., Laskar, J., Kokkotas, K., Campbell, R. M., Witzel, A. 2001, 
      A\&A, 374, 784

\bibitem[\protect\citeauthoryear{Brown et al.}{1989}]{bro89} Brown, L. M. J. et al. 1989, 
      ApJ, 340, 129

\bibitem[\protect\citeauthoryear{Capetti, Raiteri \& Buttiglione}{2010}]{cap10} Capetti, A., Raiteri, C. M., Buttiglione, S. 2010, 
      A\&A, 516, 59

\bibitem[\protect\citeauthoryear{Caproni \& Abraham}{2004a}]{caab04a} Caproni, A., Abraham, Z. 2004a, 
      ApJ, 602, 625

\bibitem[\protect\citeauthoryear{Caproni \& Abraham}{2004b}]{caab04b} Caproni, A., Abraham, Z. 2004b, 
      MNRAS, 349, 1218

\bibitem[\protect\citeauthoryear{Caproni et al.}{2006}]{cap06} Caproni, A., Livio, M., Abraham, Z., Mosquera Cuesta, H. J. 2006, 
      ApJ, 653, 112

\bibitem[\protect\citeauthoryear{Caproni, Monteiro \& Abraham}{2009}]{cap09} Caproni, A.,Monteiro, H. Abraham, Z. 2009, MNRAS, 399, 1415

\bibitem[\protect\citeauthoryear{Caproni et al.}{2011}]{cap11} Caproni, A.,Monteiro, H. Abraham, Z., Teixeira, D. M., Toffoli, R. T. 2011, ApJ, 736, 68

\bibitem[\protect\citeauthoryear{Charlot}{1990}]{char90} Charlot, P. 1990, 
      A\&A, 229, 51

\bibitem[\protect\citeauthoryear{Croke \& Gabuzda}{2008}]{crga08} Croke, S. M.,Gabuzda, D. C. 2008, MNRAS, 386, 619

\bibitem[\protect\citeauthoryear{de Boer et al}{2005}]{deb05} de Boer, P. -T., Kroese, D. P., Mannor, S., Rubinstein, R. Y.  2005, 
      Annals of Operations Research, 134, 19.	  
	  	  
\bibitem[\protect\citeauthoryear{Denn, Mutel \& Marscher}{2000}]{den00} Denn, G. R., Mutel, R. L., Marscher, A. P. 2000, 
      ApJS, 129, 61

\bibitem[\protect\citeauthoryear{Eggleton}{1983}]{eggl83} Eggleton, P. P. 1983, 
       MNRAS, 204, 449

\bibitem[\protect\citeauthoryear{Fan et al.}{1998}]{fan98} Fan, J. H., Xie, G. Z., Pecontal, E., Pecontal, A., Copin, Y. 1998, 
      ApJ, 507, 173

\bibitem[\protect\citeauthoryear{Fan, Xie \& Bacon}{1999}]{fan99} Fan, J. H., Xie, G. Z., Bacon, R. 1999, 
      A\&AS, 136, 13

\bibitem[\protect\citeauthoryear{Fasano \& Franceschini}{1987}]{fafr87} Fasano, G., Franceschini, A. 1987, 
      MNRAS, 225, 155

\bibitem[\protect\citeauthoryear{Fomalont et al.}{2000}]{fom00} Fomalont, E. B., Frey, S., Paragi, Z., Gurvits, L. I., Scott, W. K., Taylor, A. R., Edwards, P. G., Hirabayashi, H. 2000, 
      ApJS, 131, 95
	  	  
\bibitem[\protect\citeauthoryear{Gabuzda \& Cawthorne}{1996}]{gaca96} Gabuzda, D. C., Cawthorne, T. V. 1996, 
      MNRAS, 283, 759

\bibitem[\protect\citeauthoryear{Gabuzda \& Cawthorne}{2003}]{gaca03} Gabuzda, D. C., Cawthorne, T. V. 2003, 
      MNRAS, 338, 312

\bibitem[\protect\citeauthoryear{Ghisellini et al.}{2010}]{ghi10} Ghisellini, G., Tavecchio, F., Foschini, L., Ghirlanda, G., Maraschi, L., Celotti, A. 2010, 
      MNRAS, 402, 497
	  	  	  
\bibitem[\protect\citeauthoryear{Gower et al.}{1982}]{gow82} Gower, A. C., Gregory, P. C., Hutchings, J. B., Unruh, W. G. 1982, 
      ApJ, 262, 478

\bibitem[\protect\citeauthoryear{Hagen-Thorn et al.}{1997}]{hag97} Hagen-Thorn, V. A., Marchenko, S. G., Mikolaichuk, O. V., Yakovleva, V. A. 1997, Astron. Rep., 41, 154

\bibitem[\protect\citeauthoryear{Hardee}{2000}]{hard00} Hardee, P. E. 2000, 
      ApJ, 533, 176

\bibitem[\protect\citeauthoryear{Hardee}{2001}]{hard01} Hardee, P. E. 2001, 
      ApJ, 555, 744

\bibitem[\protect\citeauthoryear{Hardee}{2002}]{hard02} Hardee, P. E. 2002, 
      New Astron. Rev., 46, 427

\bibitem[\protect\citeauthoryear{Hovatta et al.}{2009}]{hov09} Hovatta, T., Valtaoja, E., Tornikoski, M., Lähteenmäki, A. 2009, 
      A\&A, 494, 527.

\bibitem[\protect\citeauthoryear{Hyv\"onen et al.}{2007}]{hyv07} Hyv\"onen, T., Kotilainen J. K., Falomo R., \"Orndahl E., Pursimo, T. 2007, 
      A\&A, 476, 723

\bibitem[\protect\citeauthoryear{Jorstad et al.}{2005}]{jor05} Jorstad, S. G., et al. 2005, AJ, 130, 1418

\bibitem[\protect\citeauthoryear{Hughes, Aller \& Aller}{1985}]{hug85} Hughes, P. A., Aller, H. D., Aller, M. F. 1985, ApJ, 289, 301

\bibitem[\protect\citeauthoryear{Hughes, Aller \& Aller}{1989}]{hug89} Hughes, P. A., Aller, H. D., Aller, M. F. 1985, ApJ, 341, 68

\bibitem[\protect\citeauthoryear{Katz et al.}{1982}]{kat82} Katz, J. I., Anderson, S. F., Margon B., Grandi, S. A. 1982, ApJ, 260, 780

\bibitem[\protect\citeauthoryear{Katz}{1997}]{katz97} Katz, J. I. 1997, ApJ, 478, 527

\bibitem[\protect\citeauthoryear{Kelly et al.}{2003}]{kel03} Kelly, B. C., Hughes, P. A., Aller, H. D., Aller, M. F. 2003, ApJ, 591, 695

\bibitem[\protect\citeauthoryear{Kovalev et al.}{2008}]{kov08} Kovalev, Y. Y., Lobanov, A. P., Pushkarev, A. B., Zensus, J. A. 2008, A\&A, 483, 759

\bibitem[\protect\citeauthoryear{Kroese, Porotsky \& Rubinstein}{2006}]{kro06} Kroese, D. P., Porotsky, S., Rubinstein, R. Y.  2006, 
      Methodology and Computing in Applied Probability, 8, 383

\bibitem[\protect\citeauthoryear{Larwood}{1997}]{larw97} Larwood, J. D. 1997, MNRAS, 290, 490

\bibitem[\protect\citeauthoryear{Lind \& Blandford}{1985}]{libl85} Lind, K. R., Blandford, R. D. 1985, ApJ, 295, 358

\bibitem[\protect\citeauthoryear{Lister et al.}{2009}]{lis09} Lister, M. L., et al. 2009, AJ, 138, 1874
	  
\bibitem[\protect\citeauthoryear{Lobanov}{1998}]{loba98} Lobanov, A. P. 1998, 
      A\&A, 330, 79

\bibitem[\protect\citeauthoryear{Marconi \& Hunt}{2003}]{mahu03} Marconi, A., Hunt, L. K. 2003, ApJ, 589, L21

\bibitem[\protect\citeauthoryear{Margolin}{2004}]{marg04} Margolin, L. 2004, 
      Annals of Operations Research, 134, 201
	  
\bibitem[\protect\citeauthoryear{Monteiro et al.}{2010}]{mon10} Monteiro, H., Dias, W. S., \& Caetano, T. C. 2010,
      A\&A, 516, 2

\bibitem[\protect\citeauthoryear{Monteiro \& Dias}{2011}]{modi11} Monteiro, H., Dias, W. S. 2011,
      A\&A, 530, 91

\bibitem[\protect\citeauthoryear{Mutel et al.}{1990}]{mut90} Mutel, R. L., Phillips, R. B., Su, B., Bucciferro, R. R. 1990, 
      ApJ, 352, 81

\bibitem[\protect\citeauthoryear{Mutel \& Denn}{2005}]{mude05} Mutel, R. L., \& Denn, G. R. 2005, 
      ApJ, 623, 79

\bibitem[\protect\citeauthoryear{O'Sullivan \& Gabuzda}{2009}]{osga09} O'Sullivan, S. P., Gabuzda, D. C. 2009, 
      MNRAS, 393, 429

\bibitem[\protect\citeauthoryear{Papaloizou \& Terquem}{1995}]{pate95} Papaloizou, J. C. B., Terquem, C. 1995, MNRAS, 274, 987

\bibitem[\protect\citeauthoryear{Pauliny-Toth \& Kellermann}{1966}]{pake66} Pauliny-Toth, I. I. K., Kellermann, K. I. 1966, ApJ, 146, 634

\bibitem[\protect\citeauthoryear{Peacock}{1983}]{peac83} Peacock, J. A., Readhead, A. C. S. 1988, 
      MNRAS, 202, 615
  
\bibitem[\protect\citeauthoryear{Pearson \& Readhead}{1988}]{pere88} Pearson, T. J., Readhead, A. C. S. 1988, 
      ApJ, 328, 114
	  
\bibitem[\protect\citeauthoryear{Polatidis et al.}{1995}]{pol95} Polatidis, A. G., Wilkinson, P. N., Xu, W., Readhead, A. C. S., Pearson, T. J., Taylor, G. B., Vermeulen, R. C., 1995, 
      ApJS, 98, 1

\bibitem[\protect\citeauthoryear{Press et al.}{1997}]{pre97} Press,W., Teukolski, S., Vetterling, W., Flannery,B., 1997, 
      Numerical recipes in C: The Art of Scientific Computing. Cambridge Univ. Press, Cambridge, 645

\bibitem[\protect\citeauthoryear{Raiteri et al.}{2010}]{rai10} Raiteri, C. M., et al. 2010, 
      A\&A, 524, 43

\bibitem[\protect\citeauthoryear{Romero et al.}{2000}]{rom00} Romero, G. E., Chajet, L., Abraham, Z., Fan, J. H. 2000, 
      A\&A, 360, 57

\bibitem[\protect\citeauthoryear{Rubinstein}{1997}]{rubi97} Rubinstein, R. Y. 1997, 
      European Journal of Operational Research, 99, 89

\bibitem[\protect\citeauthoryear{Rubinstein}{1999}]{rubi99} Rubinstein, R. Y. 1999, 
      Methodology and Computing in Applied Probability, 2, 127

\bibitem[\protect\citeauthoryear{Savolainen et al.}{2010}]{sav10} Savolainen, T., Homan, D. C., Hovatta, T., Kadler, M., Kovalev, Y. Y., Lister, M. L., Ros, E., Zensus, J. A. 2010, 
      A\&A, 512, 24

\bibitem[\protect\citeauthoryear{Shapiro \& Teukolsky}{1983}]{shte83} Shapiro, S. L., Teukolsky, S. A. 1983, in Black Holes, White Dwarfs and Neutron Stars, John Wiley \& Sons, New York, 476.

\bibitem[\protect\citeauthoryear{Stirling et al.}{2003}]{sti03} Stirling, A. M., Cawthorne, T. V., Stevens, J. A., Jorstad, S. G., Marscher, A. P., Lister, M. L., G\'omez, J. L., Smith, P. S., et al. 2003, 
      MNRAS, 341, 405
	  
\bibitem[\protect\citeauthoryear{Tateyama et al.}{1998}]{tat98} Tateyama, C. E., Kingham, K. A., Kaufmann, P., Piner, B. G., de Lucena, A. M. P., Botti, L. C. L., 1998, ApJ, 500, 810

\bibitem[\protect\citeauthoryear{Tateyama}{2009}]{tate09} Tateyama, C. E., 2009, ApJ, 705, 877
	  
\bibitem[\protect\citeauthoryear{Vermeulen et al.}{1995}]{ver95} Vermeulen, R. C., Ogle, P. M., Tran, H. D., Browne, I. W. A., Cohen, M. H., Readhead, A. C. S., Taylor, G. B., Goodrich, R. W. 1995, 
      ApJ, 452, L5

\bibitem[\protect\citeauthoryear{Villata et al.}{2004}]{vil04} Villata, M. et al. 2004, A\&A, 424, 497

\bibitem[\protect\citeauthoryear{Webb et al.}{1988}]{web88} Webb, J. R., Smith, A. G., Leacock, R. J., Fitzgibbons, G. L., Gombola,
P. P., Sheppard, D. W. 1988, AJ, 95, 374

\bibitem[\protect\citeauthoryear{Wild et al.}{2005}]{wil05} Wild, V., Peacock, J. A., Lahav, O., Conway, E., Maddox, S.,; Baldry, I. K., Baugh, C. M., Bland-Hawthorn, J., Bridges, T., Cannon, R., et al. 2005, 
      MNRAS, 356, 247

\bibitem[\protect\citeauthoryear{Wu \& Urry}{2002}]{wuur02} Woo, J.-H., Urry, C. M. 2002, ApJ, 579, 530

\bibitem[\protect\citeauthoryear{Wu et al.}{2011}]{wu11} Wu, Q., Zou, Y., Cao, X., Wang, D.; Chen, L. 2011, 
      ApJ, 740, L21

\end{thebibliography}
\end{document}